\documentclass{article}
\usepackage[utf8]{inputenc}
\usepackage{amsmath,amsthm,amssymb,multicol,xcolor}
\usepackage{graphicx}
\usepackage[margin=25truemm]{geometry}
\usepackage{istgame}
\usepackage{subcaption}
\usepackage{booktabs}
\usepackage{float}
\usepackage{array}
\usepackage{multirow}
\usepackage{mathtools}
\usepackage{authblk}
\usepackage{rotating}
\usepackage{makecell}
\usepackage{tabularray}
\usepackage[style=numeric-comp,giveninits=true,sorting=none]{biblatex}
\theoremstyle{plain}
\newtheorem*{theorem*}{Theorem}

\bibliography{ref}
\renewbibmacro{in:}{%
  \ifentrytype{article}
    {}
    {\bibstring{in}%
     \printunit{\intitlepunct}}}
\DeclareFieldFormat[article]{pages}{#1}
\DeclareFieldFormat[book]{title}{#1}
\DeclareFieldFormat[inbook]{title}{#1}
\AtEveryBibitem{%
  \clearfield{url}%
  \clearfield{isbn}%
  \clearfield{eprint}%
  \clearfield{month}%
  \clearfield{issn}%
  \clearfield{issue}%
  \clearfield{number}%
  \clearfield{doi}%
  }
\DeclareNameAlias{author}{family-given}
\renewcommand{\thesubsection}{\arabic{subsection}}

\title{Social learning with complex contagion}
\author[*]{Hiroaki Chiba-Okabe}
\author[*]{Joshua B.~Plotkin}
\affil[*]{University of Pennsylvania}
\date{}

\begin{document}

\maketitle

\begin{abstract}
\noindent  Traditional models of social learning by imitation are based on {\it{simple}} contagion --- where an individual may imitate a more successful neighbor following a single interaction. But real-world contagion processes are often complex, meaning that multiple exposures may be required before an individual considers changing their type.
We introduce a framework that combines the concepts simple payoff-biased imitation with complex contagion, to describe how social behaviors spread through a population.
We formulate this model as a discrete time and state stochastic process in a finite population, and we derive its continuum limit as an ordinary differential equation that generalizes the replicator equation, a widely used dynamical model in evolutionary game theory. When applied to linear frequency-dependent games, social learning with complex contagion produces qualitatively different outcomes than traditional imitation dynamics: it can shift the Prisoner's Dilemma from a unique all-defector equilibrium to either a stable mixture of cooperators and defectors in the population, or a bistable system; it changes the Snowdrift game from a single to a bistable equilibrium; and it can alter the Coordination game from bistability at the boundaries to two internal equilibria. The long-term outcome depends on the balance between the complexity of the contagion process and the strength of selection that biases imitation towards more successful types.
Our analysis intercalates the fields of evolutionary game theory with complex contagions, and it provides a synthetic framework to describe more realistic forms of behavioral change in social systems.
\end{abstract}

\section*{Introduction}

Social behaviors often spread by imitation, as individuals tend to imitate others they perceive as more successful.
Traditional models of this imitation process, called social learning, assume {\it{simple}} contagion: an individual will consider imitating a more successful neighbor following a single social interaction \cite{taylor1978,schuster1983replicator,boyd1988culture,,szabo1998,hofbauer2003,traulsen2007,traulsen2010,fu2011}. However, many real-world contagion processes are known to be {\it{complex}}, meaning that an individual may require multiple exposures to an alternative type before they consider changing their own type \cite{centola2018how,guilbeault2018}. Ideally, a realistic model of behavioral evolution in a population will incorporate both the (rational) tendency to imitate more successful neighbors, as well as the (irrational) tendency towards herd behavior, or conformity, that disregards material payoffs.

Complex contagion is already well developed in the study of neutral traits that spread through a population -- notably opinions or beliefs \cite{centola2007,romero2007,bakshy2012,centola2010,jalili2017,tornberg2018,iacopini2019,vasconcelos2019} -- when there is no intrinsic benefit of adopting one type or another.  And yet there are many real-world situations where social behaviors with frequency-dependent payoffs, such as cooperation \cite{suri2011,fowler2010} or adoption of online services with network externalities \cite{ugander2012, karsai2014,karsai2016}, also spread through social contagion. And so it is natural to postulate that both herd mentality, embodied in complex contagion, as well as payoff-biased social learning, captured by imitation dynamics, collectively drive behavioral evolution. 

Of particular interest is the interplay between these two qualitatively different aspects of behavioral change -- because the need for multiple exposures may arise from considerations outside of the direct material payoff of the behavior itself, such as credibility, legitimacy, and emotion \cite{centola2007}. Although theoretical studies of complex contagion have focused on different network structures \cite{guilbeault2018}, the question of how complex contagion modifies payoff-biased imitation has received less attention. 

Here, we propose a mathematical model that combines the dynamics of complex contagion with those of payoff-biased imitation. In this model, individuals interact with a subset of their peers in the population, compare their fitness to others, and perform payoff-biased imitation. The imitation process involves complexity in the sense that an individual considers switching types only if the frequency of the alternative type encountered exceeds some threshold. The threshold frequency for contagion is drawn independently for each instance of a possible imitation event, from a fixed power-law distribution that depends on a complexity parameter $\gamma$. We derive an ordinary differential equation (ODE) that approximates the dynamics of this discrete time/state stochastic model, in the limit of both a large population size and large neighborhood size of a focal individual. The resulting ODE reduces to the replicator equation, the classic model of payoff-biased social learning  \cite{taylor1978,schuster1983replicator,boyd1988culture,traulsen2006,cressman2014}, for simple contagions ($\gamma=1$). When contagion is complex ($\gamma \ne 1$) we analyze the long-term behavioral outcomes of payoff-biased imitation, focusing on linear games. 

Our analysis reveals that the presence and nature of complex contagion can qualitatively change long-term behavior in a population, compared to conventional predictions of evolutionary game theory. In the Prisoner's Dilemma, which is traditionally characterized by a unique stable equilibrium with everyone defecting, certain types of contagion complexity can shift this equilibrium to an interior equilibrium with a mixture of both types, or alternatively to a bistable systems with equilibria at each boundary. The classic Snowdrift game may become bistable instead of having a unique stable interior equilibrium; the Coordination game, typically characterized by bistability with the population  adopting one or the other strategy, can instead have two different internal stable equilibria. We conclude by discussing the relevance of these results, and our mathematical formulation, for more realistic models of social learning that combine both rational, payoff-biased updates along with features of conformity or anti-conformity biases \cite{asch2016effects,richerson2008not,
efferson2008conformists,
whalen2015conformity,
denton2020cultural,
newberry2022measuring}.

\section*{Model}

We introduce features of payoff-biased imitation into a model of neutral complex contagion.  \cite{vasconcelos2019} defined a neutral complex contagion in a structured population, where  each individual $i=1,2,\cdots N$ is represented by a vertex on a network. The population consists of two types $X,Y$, and individuals change their type based on a complex contagion among neighbors. Specifically, at each time step, a randomly chosen focal individual $i$ switches its type with some probability. If the focal individual is of type $X$, it switches to type $Y$ with probability

\begin{equation*}
p^{X\to Y}_{i}=y_{i}^{\gamma}
\end{equation*}
where $y_{i}$ is the proportion of neighbors of $i$ of type $Y$; otherwise individual $i$ retains his type. If the focal individual is of type $Y$, the probability of   switching to type $X$ is symmetrically defined as $p_{i}^{Y\to X}=x^{\gamma}_{i}$. Although \cite{vasconcelos2019} considered a more general situation where $\gamma$ differs between $p_{i}^{X\to Y}$ and $p_{i}^{Y\to X}$, we fix $\gamma$ to be identical for both directions, and we focus instead on asymmetries in switching probabilities that arise from payoff differences between types, which will in general depend on their frequencies in the population. 

The stochastic update rule above can be interpreted mechanistically as follows: individual $i$ has some threshold, $M$, that compels them to switch types. Individual $i$ will switch only if they have $M$ or more neighbors of the other type. The threshold $M$ is  drawn independently by each individual, from an identical probability distribution $d(M)\propto M^{\gamma - 1}$ \cite{vasconcelos2019}. With a normalizing constant $\gamma$, we obtain the desired switching probability. For example, when $i$ is of type $X$,

\begin{equation*}
\begin{split}
p_{i}^{X\to Y}&=P(y_{i}\geq M|y_{i})\\
&=\int_{0}^{y_{i}}\gamma M^{\gamma-1}dM\\
&=y_{i}^{\gamma}.
\end{split}
\end{equation*}
When $\gamma=1$ any value of threshold $M\in [0,1]$ is drawn with equal chance, so that the switching probability equals the frequency of the opposite type in an individual's neighborhood. This corresponds to a simple contagion. A value $\gamma >1$ means that an individual is more likely to have a high threshold for switching types, whereas $\gamma < 1$ means a lower threshold is more likely. This model of complex contagion describes the spread of 
{\it{neutral}} markers $X$ and $Y$, in the sense that an individual's type does not influence fitness or bias the process of imitation. Neutral models have been widely used to study the contagion of opinions or beliefs \cite{centola2007,romero2007,bakshy2012,centola2010,jalili2017,tornberg2018,iacopini2019,vasconcelos2019}.  

We will generalize this complex contagion process, focusing on a well-mixed population of individuals who engage in pairwise social interactions. In our setting, an individual's type represents a strategy for social interaction, and individuals acquire payoffs that depend on their own strategy and the strategies of those they interact with. We seek to generalize the model of contagion above so that it is biased by payoffs -- meaning that individuals tend to copy strategies that are more successful -- while retaining the threshold-like behavior of a complex contagion \cite{granovetter1978,morris2000,watts2002,centola2007,jalili2017}. To develop this combined model, we stipulate that in each discrete time period, each individual plays a pairwise game with $n$ other individuals, randomly selected from the population (and possibly including self-play). At the end of each time period, one individual $i$ is selected to consider updating their strategy. Let $n_{X}$ be the number type-$X$ individuals the focal individual $i$ interacted with, and $n_{Y}$  the number of type-$Y$ individuals in that time period ($n=n_{X}+n_{Y}$). We posit that a focal individual $i$ of type $X$ will then switch to strategy $Y$ with probability $\rho^{X\to Y}$ (and likewise a focal individual of type $Y$ will switch to type $X$ with probability $\rho^{Y\to X}$) defined by the following expression:
\begin{equation}
\rho^{X\to Y}=
\left\{
\begin{array}{ll}
\frac{1}{1+\exp(sD)} & \textrm{if } n_{Y}/n \geq M \\
0 & \textrm{otherwise}
\end{array}
\right. \ \ \ \ \ \ \ 
\rho^{Y\to X}=
\left\{
\begin{array}{ll}
\frac{1}{1+\exp(-sD)} & \textrm{if } n_{X}/n \geq M \\
0 & \textrm{otherwise}
\end{array}
\right.
\label{eq:rho}
\end{equation}
where $D\triangleq f_{X}(n_{X})-f_{Y}(n_{X})$ is the difference in total payoffs to individuals of types $X$ and $Y$ in that time period. 
The fitness of each type is the empirical payoff perceived by the focal individual $i$: $f_{X}(n_{X})=[n_{X} \pi(X,X)+n_{Y}\pi(X,Y)]/n, 
f_{Y}(n_{X})=\left[n_{X}\pi(Y,X)+n_{Y} \pi(Y,Y)\right]/n$ where $\pi(X,Y)$ denotes the payoff a type-$X$ individual playing the pairwise game with an opponent of type $Y$, and so on.

Our definition of the switching probability (Eq.~\ref{eq:rho}) involves a Fermi function of the payoff difference between individuals, which is a standard formulation for payoff-biased imitation dynamics \cite{hauert2005,traulsen2006,traulsen2007,jusup2022}. The parameter $s$ denotes the strength of selection \cite{traulsen2009} that quantifies how much payoffs matter at all when individuals consider switching types. At the same time, the switching probability in Eq.~\ref{eq:rho} also involves a threshold frequency $M$, characteristic of complex contagion. And so this model is a combination of payoff-biased imitation along with complex contagion.

\section*{Results}

\subsection{A replicator equation for complex contagions}

We will analyze this model of social learning with complex contagion by studying its continuum limit, derived for large populations. We focus on well-mixed population for which analytical expression of the dynamics can be derived. Specifically, as both $N$ and $n$ becomes large, the dynamics can be approximated by the following ODE (see Methods for derivation).
\begin{equation}
\label{eq:complexreplicator}
\dot{x}=x(1-x)\left(x^{\gamma-1}(1+\alpha D)-(1-x)^{\gamma-1}(1-\alpha D)\right)
\end{equation}
where $x$ denotes the frequency of strategy $X$ (as opposed to strategy $Y$) in the population, and $\alpha\triangleq s/2$ reflects the strength of selection.

This model provides a sound basis for studying complex contagion of strategic behaviors with payoff-biased imitation in large populations. When $\gamma=1$, Eq. \ref{eq:complexreplicator} reduces to the standard replicator equation for strategy dynamics under simple contagion. On the other hand, in the absence of selection, $\alpha=0$, this model reduces to the standard model of complex contagion for neutral types. All other regimes --- that is, the complex contagion of behavioral strategies that affect fitness and imitation --- remain unexplored.

We note in passing that there is an alternative construction of this model from a discrete stochastic process that also gives rise to these replicator dynamics in two-strategy settings (see Methods).

\subsection{Fixed points and their stability}

Our model for the dynamics of complex, payoff-biased imitation (Eq.~\ref{eq:complexreplicator}) has two trivial equilibria, $x=0$ and $x=1$, as well as possible internal equillibria that can be obtained by solving

\begin{equation*}
\bar{D}(x):=x^{\gamma-1}(1+\alpha D)-(1-x)^{\gamma-1}(1-\alpha D)=0.
\end{equation*}

The quantity $\bar{D}(x)$ represents an appropriately scaled version of the payoff difference $D(x)$ that accounts for the complexity parameter $\gamma$. For linear games, where the payoff difference $D$ is a linear function of the strategy frequency $x$, we can analytically determine the number of internal fixed points (i.e.~solutions to the above equation) for several different choices of $\gamma$ and $\alpha$ (see Methods for details). In particular, if $\alpha=0$ (e.g.~a Donation game), there is at most  one internal fixed point regardless of the sign of $\gamma$. When $\alpha > 0$ (e.g.~Coordination game), there is at most one internal fixed point if $\gamma > 1$. When $\alpha < 0$, there is at most one internal fixed point if $\gamma < 1$.

The stability at a fixed point is determined by the sign of $\bar{D}$. Specifically, $x=0$ is an asymptotically stable steady state if $\lim_{x\rightarrow 0}\bar{D}<0$; and $x=1$ is an asymptotically stable steady state if $\lim_{x\rightarrow 1}\bar{D}>0$. Furthermore, since that the pre-factor $x(1-x)$ in the continuous dynamics is strictly positive and bounded in the interior, there is a stable (unstable) equilibrium whenever $\bar{D}$ downcrosses (upcrosses) the $x$-axis in the interior of $[0,1]$ (see Methods). This simple condition allows us to analyze all equilibria and their stability properties simply by  plotting $\bar{D}(x)$ as a function of the frequency $x$ of strategic type $X$.

\subsection{Behavioral evolution for linear games}

The framework above allows us to characterize the long-term outcomes of payoff-biased complex contagion of behavioral strategies, for any choice of contagion parameter $\gamma$ and any $2 \times 2$ symmetric game matrix {\tiny$\setlength\arraycolsep{1pt}\begin{pmatrix} R & S \\ T & P \end{pmatrix}$}. The following examples show how complex contagion can lead to qualitatively different outcomes for which behaviors dominate, compared to the standard model of simple payoff-biased imitation. For example, the Prisoner's Dilemma always leads to a population of pure defectors for simple contagion of behaviors; but under complex contagion with $\gamma<1$ the population can evolve to a stable mixture of cooperators and defectors. We will also analyze Snowdrift and Coordination games.

\begin{figure}[!ht]
    \centering
    \includegraphics[width=\linewidth]{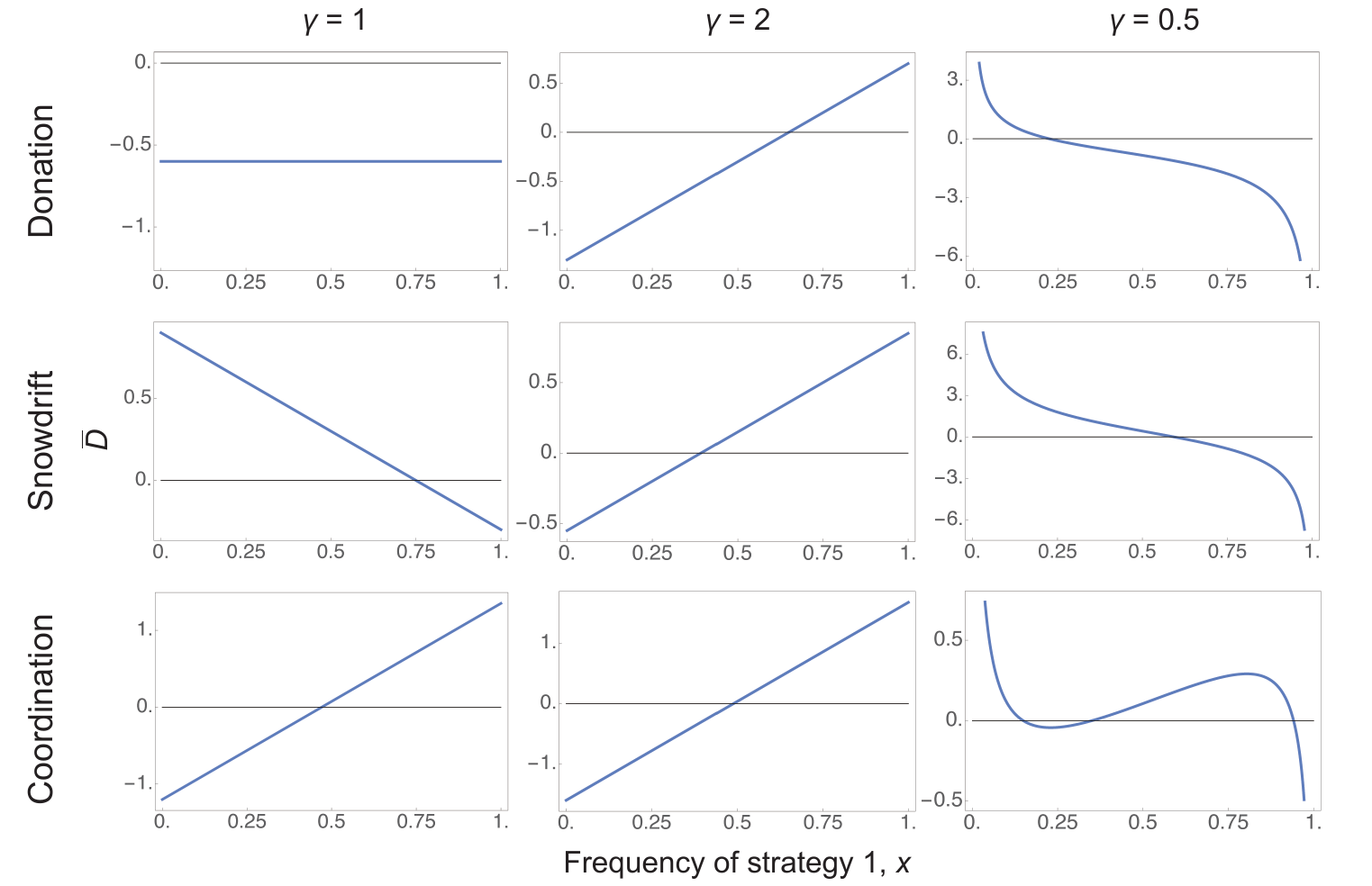}
    \caption{\small \textbf{Complex contagion can qualitatively change outcomes of behavioral evolution.}
    We plot the scaled payoff difference $\bar{D}(x)$ as a function of the frequency $x$ of strategy 1, for three different games. A stable internal equilibrium, where both strategies co-exist, occurs when $\bar{D}$ intersects the $x$-axis with negative slope. The equilibrium at $x=0$ ($x=1$) is stable if $\bar{D}$ is negative (positive) at that boundary. Under simple contagion ($\gamma=1$), the only stable outcomes are pure defection in the Prisoner's Dilemma, a mixture of both strategies in the Snowdrift game, and pure coordination in the Coordination game. But complex contagion ($\gamma>1$ or $\gamma<1$) leads to qualitatively different outcomes, including, e.g., a stable mixture of cooperators and defectors in a Prisoner's dilemma. Parameters: $\alpha=0.15$, $b=5$, $c=2$, $p=4.5$, $q=4$.
    \label{fig:complex}}
\end{figure}

Figure~\ref{fig:complex} plots the difference in payoffs of the two strategies, $\bar{D}$, as a function of the frequency $x$ of one strategy. 
These plots graphically depict the possible dynamical outcomes, shown for the Donation game, Snowdrift game, and Coordination game. An internal equilibrium exists whenever $\bar{D}$ intersects the $x$-axis, and it is stable (unstable) if it is down-crossing (up-crossing). The boundary equilibrium at $x=0$ ($x=1$) is stable if $\bar{D}$ is negative (positive) at that point. 

We see that a contagion with $\gamma<1$, which is form of anti-conformity bias, tends to create a stable internal equilibrium, because this type of contagion helps permit a strategy with lower payoff to spread. Conversely, conformity bias $\gamma>1$ tends to create equilibria at endpoints, because this contagion process  makes individuals more likely to cluster on one strategy or the other. The effects of the contagion process in both cases, however, are also moderated by the dynamics arising payoffs, to a degree that depends on the strength of selection $\alpha$ in payoff-biased imitation.

We specify the payoff matrices for each of these three games (Prisoner's Dilemma, Snowdrift, and Coordination) below. For each game we provide provide explicit expressions for the internal equilibria, for several different values of the parameter $\gamma$ that permit closed-form analytical solutions. A comprehensive analysis of the dynamics of these games under all parameter regimes is shown in Table~\ref{tab:linear games}, and a more general treatment of all linear games is described in Methods. 

\paragraph{Donation Game}
The Donation game is a minimal model a Prisoner's Dilemma with payoff matrix as follows:
\begin{equation*}
    \begin{pmatrix}
        b-c & -c \\
        b & 0
    \end{pmatrix}
\end{equation*}
where \(b > c > 0\). For $\gamma=1/2$ and $\gamma=2$, we can explicitly solve for the unique internal equilibrium, $x^{\ast}=\frac{(1-\alpha c)^{2}}{2(1+\alpha^{2} c^{2})}$ and $x^{\ast}=\frac{1+\alpha c}{2}$, respectively.

\paragraph{Snowdrift Game}
A typical Snowdrift game has the following payoff matrix:
\begin{equation*}
    \begin{pmatrix}
        b - \frac{c}{2} & b - c \\
        b & 0
    \end{pmatrix}
\end{equation*}
where \(b > c > 0\). For $\gamma=2$, we can explicitly solve for the unique internal equilibrium $x^{\ast}=\frac{2(1-b+c)}{4-\alpha(2b-c)}$.

\paragraph{Coordination Game}
A typical Coordination game has payoff matrix 
\begin{equation*}
    \begin{pmatrix}
        p & 0 \\
        0 & q
    \end{pmatrix}
\end{equation*}
where \(p > q > 0\). For $\gamma=2$, we can explicitly solve for the unique internal equilibrium $x^{\ast}=\frac{1+\alpha q}{2+\alpha(p+q)}$.

The examples above illustrate that analytical solutions are possible for the internal equillibria of these three classical games, for specific values of the contagion parameter $\gamma$. More generally, Table 1 summarizes a full analysis of all qualitative outcomes, as a function of the contagion parameter, the game payoff matrix, and the strength of payoff-biased imitation. In particular, for each of these three classic games, Table 1 specifies what parameter regimes will lead to different long-term behavioral outcomes -- i.e., dominance of one strategy, bi-stability, or a stable mixture of both strategies.

\vspace{5mm}

\begin{table}[!ht]
\centering
\begin{tabular}{cccccc}
&&&\\
\toprule
\multirow{2}{*}{Game} & \small{Complexity}  &  \small{Game and Imitation} & \multirow{2}{*}{$x = 0$} & \multirow{2}{*}{$x \in (0,1)$} &  \multirow{2}{*}{$x = 1$} \\
&\small{Parameter}& \small{Parameters} & \multicolumn{3}{c}{} \\
\midrule \multirow{5}{*}{Donation}
                          & $\gamma=1$ & all& stable & none & unstable \\
 & \multirow{2}{*}{$\gamma > 1$ } &  $c<1/\alpha$ & stable & unstable \& unique & stable \\
                          &                              &  $1/\alpha < c$ & stable & none & unstable \\

                          & \multirow{2}{*}{$\gamma < 1$} &  $c<1/\alpha$ & unstable &  stable \& unique & unstable \\
                          &                              &  $1/\alpha<c$ & stable &  none & unstable \\
\midrule
\multirow{5}{*}{Snowdrift} 
                           & $\gamma\leq 1$ & all & unstable & stable \& unique & unstable \\
                            & \multirow{4}{*}{$\gamma > 1$} &  $c < 2/\alpha$, $1/\alpha < b-c$ & unstable & indeterminate & stable \\
                           &                              & $2/\alpha < c$, $1/\alpha < b-c$ & unstable & stable \& potentially non-unique & unstable \\
                           &                              & $c < 2/\alpha$, $b-c < 1/\alpha$ & stable & unstable \& potentially non-unique & stable \\
                           &                              & $2/\alpha < c$, $b-c < 1/\alpha$ & stable & none & stable \\

\midrule
\multirow{3}{*}{Coordination} & $\gamma\geq 1$ & all & stable & unstable \& unique & stable \\
                                 & \multirow{2}{*}{$\gamma < 1$} & $q<1/\alpha<p$ & unstable & indeterminate & stable \\
                                 &                              & $p<1/\alpha$ & unstable & stable \& potentially non-unique& unstable \\
\bottomrule
\end{tabular}
\caption{\small \textbf{Fixed points and their stability for three linear games under payoff-biased complex contagion.} Complex contagion changes the equilibria of the standard replicator equation ($\gamma=1$).  The long-term behavioral outcome -- dominance or one strategy or the other,  $x=0$ or $x=1$, bi-stability, or a stable mixture of strategies $x \in (0,1)$ -- depends on the complexity parameter $\gamma$ as well as the game payoffs and the strength of payoff-biased imitation, $\alpha$.}
\label{tab:linear games}
\end{table}

\subsection{Monte Carlo simulations of the stochastic process}

We undertook stochastic simulation of the underlying discrete time/state stochastic process, to compare their behavior to the ODE derived in the infinite-population limit.
We set the population size $N=2,500$, neighborhood size $n=500$,  and we ran Monte Carlo simulations for $10^{6}$ time steps with different initial strategic frequencies. These simulations (Table S1)  show good agreement with the continuum ODE approximation.\footnote{Computer code for the simulations can be found on GitHub: \url{https://github.com/hirochok/slcc}.}

\subsection{Games with more than two strategies}

We can generalize our modelling framework for complex contagion and payoff-biased imitation to accommodate games with more than two strategies. Suppose there are $m$ different strategies $X_{1},X_{2},\cdots,X_{m}$ whose frequencies in the population are denoted  $x_{1},x_{2},\cdots,x_{m}$. In order to generalize our modelling framework for $m>2$ strategies we define $\rho$ as

\begin{equation*}
\begin{split}
\rho^{X_{j}\to X_{k}}&=
\left\{
\begin{array}{ll}
\frac{\exp\left(sf_{X_{k}}(\vec{n})\right)}{\sum_{l}\exp\left(sf_{X_{l}}(\vec{n})\right)} & \textrm{if } n_{X_{k}}/n \geq M\\
0 & \textrm{otherwise}
\end{array}
\right.\\
&=
\left\{
\begin{array}{ll}
\frac{1}{1+\sum_{l}\exp\left[s\left(f_{X_{l}}(\vec{n})-f_{X_{k}}(\vec{n})\right)\right]} & \textrm{if } n_{X_{k}}/n \geq M\\
0 & \textrm{otherwise}
\end{array}
\right.
\end{split}
\end{equation*}
where $n_{X_{j}}$ is the number of type-$X_{j}$ individuals with whom a focal individual interacted in each time step, $\vec{n}\triangleq (n_{X_{1}},\cdots,n_{X_{m}})$, and the fitness of an arbitrary type $j$ is defined as the empirical payoff perceived by the focal individual: $f_{X_{j}}(\vec{n})=\sum_{l}n_{X_{l}}\pi(X_{j},X_{l})/n$.

This definition for $\rho$ reduces to Eq.~\ref{eq:rho} in the case of $m=2$ strategies. And so this framework is a natural generalization of the two-strategy case. Also, note that $\sum_{k}\left[\exp\left(sf_{X_{k}}(\vec{n})\right)/\sum_{l}\exp\left(sf_{X_{l}}(\vec{n})\right)\right]=1$, so that the probability of switching to another type never exceeds $1$, even if the threshold $M$ is reached for multiple other types, thus preserving the total mass of the population.

In the limit of $N\rightarrow \infty$, we can approximate the behavior of this stochastic discrete process by a system of ODEs that reduces to the classical replicator equation in the case of simple contagion ($\gamma=1$, see Methods):

\begin{equation}
\label{eq:m-type}
\dot{x}_{j}=\sum_{k}x_{k}x_{j}^{\gamma}\left[\frac{1}{m}-\frac{s}{m^{2}}\sum_{l\neq j}\left(f_{X_{l}}(\vec{x})-f_{X_{j}}(\vec{x})\right)\right]-x_{j}\sum_{k}x_{k}^{\gamma}\left[\frac{1}{m}-\frac{s}{m^{2}}\sum_{l \neq k}\left(f_{X_{l}}(\vec{x})-f_{X_{k}}(\vec{x})\right)\right]
\end{equation}
where $\vec{x}\triangleq (x_{1},\cdots,x_{m-1})$. 

We can use Eq.~\ref{eq:m-type} study how complex contagion alters strategy dynamics in multi-choice games. We focus on the interesting case of the non-transitive Rock-Paper-Scissor game with payoff matrix

\begin{equation*}
\begin{pmatrix}
0&-1&1\\
1&0&-1\\
-1&1&0
\end{pmatrix}.
\label{eq:RPS}
\end{equation*}

\begin{figure}[!ht]
\centering
\includegraphics[width=.9 \linewidth]{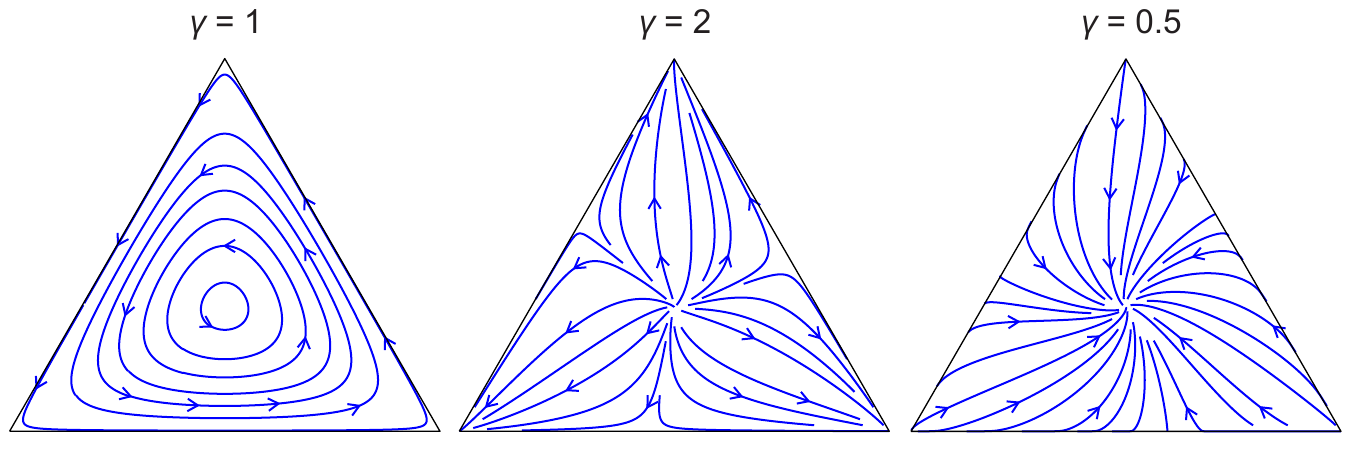}
\caption{\small \textbf{Strategy evolution in the Rock-Paper-Scissor game with simple and complex contagion.} 
Each vertex represents the monomorphic population of each strategy. When $\gamma=1$ (simple contagion and classic replicator equation), the dynamics exhibit a cyclic nature. Whereas the population is attracted to each monomorphic state when $\gamma>1$, and it attracted to a stable mixture of strategies when $\gamma<1$. Parameters: $s=0.3$.}
\label{fig:RPS}
\end{figure}

When $\gamma=1$ (simple contagion) the replicator dynamics are already well understood, and they exhibit a cyclic pattern. For $\gamma > 1$, however, the dynamics gravitate towards the three pure strategy states, each of which is stable (Figure~\ref{fig:RPS}). Whereas for  $\gamma < 1$
there is a single stable equilibrium with a mixture of all three strategies present in the population (Figure~\ref{fig:RPS}). We also use this ODE to derive results for a three-strategy generalization of the Prisoner's Dilemma game (see Supplementary Material), where again we find that complex contagion qualitatively changes dynamical outcomes.

\section*{Discussion}

This study takes a first step towards integrating  complex contagions with payoff-biased imitation to describe the evolution of social traits in a population. We have developed a framework that incorporates elements of both of these processes simultaneously, which extends the literature on payoff-biased {\it{simple}} contagions \cite{taylor1978,schuster1983replicator,boyd1988culture,helbing1992,hofbauer2003,bauch2005,traulsen2007,traulsen2010,fu2011} as well as the literature on complex contagions for neutral traits  \cite{centola2007,jalili2017,tornberg2018,iacopini2019,vasconcelos2019}. The resulting synthetic framework offers a way to study realistic situations where behavioral change is driven in part by aspects of conformity or anti-conformity biases 
\cite{asch2016effects,richerson2008not,
efferson2008conformists,
whalen2015conformity,
denton2020cultural,
newberry2022measuring} and in part by a tendency to imitate more successful types \cite{taylor1978,schuster1983replicator,hofbauer2003,bauch2005,traulsen2007,traulsen2010,fu2011,mbah2012,allen2017}. While this framework can be seen as a generalization of both payoff-biased imitation and complex contagion dynamics, it is particularly suited for describing scenarios where both elements are present \cite{fowler2010,suri2011,ugander2012,karsai2014,karsai2016}.

Our analysis reveals that the complexity of the contagion process can qualitatively change the long-term behavioral outcomes in a population undergoing social learning -- e.g., changing the eventual dominance of one type into either a stable mixture of two types, or alternatively producing bistability where either one or the other type will eventually dominate thepopulation.

Our modelling framework is developed as a discrete time and state stochastic process that described when one individual will imitate another's type. Taking the limit as the population size gets large we derived a replicator-like ODE, which seamlessly connects standard complex contagions of neutral types \cite{vasconcelos2019} and the standard replicator equation for payoff-based strategy evolution\cite{taylor1978}. The resulting ODE, whose behavior matches the quasi-stationary behavior of stochastic simulations in a finite population, provides a tool for analyzing the dynamics of strategies for linear pairwise games, with two or more strategies. Furthermore, it offers a novel interpretation of the standard replicator equation: it describes a dynamic that arise when individuals engage in payoff-biased imitation only if the frequency of an alternative strategy is beyond certain threshold that is drawn from a uniform distribution ($\gamma=1$). 

The introduction of complexity into the contagion processes ($\gamma \ne 1$) qualitatively changes long-term behavior in  evolutionary games. In the Prisoner’s Dilemma, complexity can shift the system from a unique monomorphic stable state (defection) to either an interior equilibrium (a mixture of cooperators and defectors) or to a bistable state (defection or cooperation, depending upon initial conditions). The Snowdrift game, under complex contagion, transitions from a unique stable interior equilibrium to exhibiting bistability. Lastly, the Coordination game can shift from its usual bistable nature to having two internal stable equilibria. The parameter $\gamma$, which governs the power-law distribution from which the threshold for considering alternative strategies is drawn, and the parameter $\alpha$, which determines the strength of payoff-biased imitation, collectively determine the long-term evolutionary outcomes.

This study sheds light on the interplay between complex contagion and payoff-biased imitation, but it is certainly not without limitations. First, we have focused on the case where the population is well-mixed. Future work could extend our analysis to structured populations, either for meta-populations or for populations with an explicit network structure of pairwise social interactions. Also, we have considered independent and identical power-law distribution for the contagion threshold $M$, and we focused our analysis on linear frequency-dependent games. Different threshold distributions, perhaps dependent on an individual's strategy or even their position within a social network, as well as applications to non-linear games are all avenues for future research. Perhaps the most important question -- namely, the relative role of contagion complexity versus payoff-bias for behavioral imitation in real populations -- remains unresolved, and it will require inference of model parameters from empirical data. This study nonetheless provides an elementary framework to pose such questions and, eventually, to draw quantitative conclusions about the strength of conformity biases versus payoff biases in the evolution of social behaviors.

\clearpage
\small{
\section*{Methods}
\setcounter{subsection}{0}
\subsection{Derivation of the continuous limit ODE}

Our microscopic model describes complex contagion among strategic types as a discrete-time, discrete-state stochastic process in a finite population. The process tracks of the probability density $P^{\tau}(j)$ for the number of type-$X$ individuals, $j$, at each time point $\tau$. The transition probabilities, $T^{+}_{X}(j)$ and $T^{-}_{X}(j)$, that the number of type-$X$ individuals will increase or decrease by one each time unit when the current state is $j$ are defined by
\begin{equation*}
T^{+}_{X}(j)=\frac{N-j}{N}\sum_{k=0}^{n}\rho^{Y\to X}|_{M,n_{X}} P(n_{X}=k) \ \ \ \ \ 
T^{-}_{X}(j)=\frac{j}{N}\sum_{k=0}^{n}\rho^{X\to Y}|_{M,n_{X}}P(n_{X}=k).
\end{equation*}
We  focus on the case of a well-mixed population of size $N$, so that the composition of encounters, $P(n_{X}=k)$, is given by a binomial distribution.

The discrete process is described by its associated master equation \cite{gardiner2004,traulsen2005}:

\begin{equation*}
\begin{split}
P^{\tau+1}(j)-P^{\tau}(j)=&P^{\tau}(j-1)T_{X}^{+}(j-1)-P^{\tau}(j)T_{X}^{-}(j)\\
&+P^{\tau}(j+1)T_{X}^{-}(j+1)-P^{\tau}(j)T_{X}^{+}(j).
\end{split}
\end{equation*}

We follow the standard approach to derive an associated continuous-state/continuous-time limit, as $N\rightarrow \infty$. We track frequencies instead of counts and re-scaling time,  defining $x\triangleq j/N$, $t\triangleq \tau/N$ and $p(x,t)\triangleq NP^{\tau}(j)$ which yields the description
\begin{equation*}
\begin{split}
p(x,t+N^{-1})-p(x,t)=&p(x-N^{-1},t)T^{+}_{X}(x-N^{-1})+p(x+N^{-1},t)T^{-}_{X}(x+N^{-1})\\
&-p(x,t)T^{-}_{X}(x)-p(x,t)T^{+}_{X}(x).
\end{split}
\end{equation*}
Neglecting  higher-order terms of the Kramers-Moyal expansion, we obtain a corresponding Fokker-Planck equation
\begin{equation*}
\frac{d}{dt}p(x,t)=-\frac{d}{dx}[a(x)p(x,t)]+\frac{1}{2}\frac{d^{2}}{dx^{2}}[b^{2}(x)p(x,t)]
\end{equation*}
where $a(x)=T^{+}_{X}(x)-T^{-}_{X}(x)$ and $b(x)=\sqrt{N^{-1}(T_{X}^{+}(x)+T_{X}^{-}(x))}$. According to Ito Calculus, the process described by the Fokker-Planck equation is equivalent to a Langevin equation \cite{gardiner2004}:
$\dot{x}=a(x)dt+b(x)dW_{t}$
where $dW_{t}$ represents the increment of a Wiener process. The noise term vanishes as $N$ becomes large. 

We have assumed  $M$ is i.i.d across individuals and time, and drawn from the distribution $d(M)=\gamma M^{\gamma-1}$. Then, assuming $n$ is large enough,
\begin{equation*}
\begin{split}
T^{-}_{X}(x)&=x\sum_{k=0}^{n}\rho^{X\to Y}|_{M,n_{X}}P(n_{X}=k)\\
&=x\sum_{k=0}^{n}\frac{(1-k/n)^{\gamma}}{1+\exp(-sD(k/n))} \binom{n}{k}x^{k}(1-x)^{n-k}\\
&\approx \frac{x(1-x)^{\gamma}}{1+\exp(-sD(x))}\\
\end{split}
\end{equation*}
and similarly
\begin{equation*}
T^{+}_{X}(x)\approx \frac{(1-x)x^{\gamma}}{1+\exp(sD(x))}.
\end{equation*}
Here $D(x)$ again denotes the difference in payoffs to individuals of type $X$ and $Y$, given the current frequency $x$ of type $X$ in the population; and the parameter $s$ governs the strength of selection. 

The approximation above follows from the following property of Bernstein polynomial \cite{lai1992} (see also e.g. \cite{pena2016} for other applications of Bernstein polynomial to evolutionary dynamics).

\begin{theorem*}
Let $f$ be a continuous function on the interval $[0,1]$ and $B_{n}(f)$ be the $n$th Bernstein polynomial
\begin{equation*}
B_{n}(f):=\sum_{k=0}^{n}\binom{n}{k}x^{k}(1-x)^{n-k}f\left(\frac{k}{n}\right).
\end{equation*}
Then, $B_{n}(f)(x)$ converges uniformly to $f(x)$ as $n\rightarrow \infty$.
\end{theorem*}

For rotational simplicity we suppress the fact that $D(x)$ depends on $x$ and write $D$. Observe that

\begin{equation*}
\begin{split}
\frac{1}{1+\exp(sD)}&= \left.\frac{1}{1+\exp(sD)}\right|_{s=0}+s\left.\left(\frac{d}{ds}\left[\frac{1}{1+\exp(sD)}\right]\right)\right|_{s=0}+\mathcal{O}(s^{2})\\
&\approx \frac{1}{2}-\frac{sD}{4}.
\end{split}
\end{equation*}
Thus, we obtain the approximate continuous dynamics

\begin{equation*}
\dot{x}=(1-x)x^{\gamma}\left(\frac{1}{2}+\frac{sD}{4}\right)-x(1-x)^{\gamma}\left(\frac{1}{2}-\frac{sD}{4}\right).
\end{equation*}
With appropriate normalization, and defining $\alpha=s/2$, we arrive at the following continuous-time dynamical equation in the limit of large $N$:
\begin{equation*}
\begin{split}
\dot{x}&=x^{\gamma}y(1+\alpha D)-xy^{\gamma}(1-\alpha D)\\
\dot{y}&=y^{\gamma}x(1-\alpha D)-yx^{\gamma}(1+\alpha D).
\end{split}
\end{equation*}
Note $\dot{x}+\dot{y}=0$ and the so dynamics can be described by a single ODE
\begin{equation*}
\begin{split}
\dot{x}&=x^{\gamma}(1-x)(1+\alpha D)-x(1-x)^{\gamma}(1-\alpha D)\\
&=x(1-x)\left(x^{\gamma-1}(1+\alpha D)-(1-x)^{\gamma-1}(1-\alpha D)\right).
\end{split}
\end{equation*}

\subsection{Fixed points and stability analysis}\label{sec:fixed points and stability}

The equation above has trivial equilibria at $x=0$ and $x=1$. Internal equilibria can be obtained by solving

\begin{equation*}
\begin{split}
\bar{D}(x):=&x^{\gamma-1}(1+\alpha D)-(1-x)^{\gamma-1}(1-\alpha D)=0\\
&\Leftrightarrow \left(\frac{x}{(1-x)}\right)^{\gamma-1}=\frac{1-\alpha D}{1+\alpha D}\\
&\Leftrightarrow \frac{x}{1-x}=\left(\frac{1-\alpha D}{1+\alpha D}\right)^{\frac{1}{\gamma-1}}\\
&\Leftrightarrow \left[1+\left(\frac{1-\alpha D}{1+\alpha D}\right)^{\frac{1}{\gamma-1}}\right]x-\left(\frac{1-\alpha D}{1+\alpha D}\right)^{\frac{1}{\gamma-1}}=0
\end{split}
\end{equation*}
whose solution is given by

\begin{equation*}
x^{\ast}=\frac{\left(\frac{1-\alpha D}{1+\alpha D}\right)^{\frac{1}{\gamma-1}}}{1+\left(\frac{1-\alpha D}{1+\alpha D}\right)^{\frac{1}{\gamma-1}}}.
\end{equation*}

Although the expression for $x^{\ast}$ looks onerous, we can often determine some properties of the internal fixed points for linear games as discussed in Section~\ref{sec:linear games}. The stability at fixed points can be studied by considering the sign of $d\dot{x}/dx$. 

\begin{equation*}
\begin{split}
\frac{d\dot{x}}{dx}=&\left[\gamma x^{\gamma-1}-(1+\gamma)x^{\gamma}\right](1+\alpha D)\\
&-\left[(1-x)^{\gamma}+\gamma x(1-x)^{\gamma-1}\right](1-\alpha D)\\
&+\left[x^{\gamma}(1-x)+x(1-x)^{\gamma}\right]\alpha \frac{dD}{dx}
\end{split}
\end{equation*}
At the boundaries of the state space, for bounded $D$, we have

\begin{equation*}
\begin{split}
\underset{x\rightarrow 0}{\lim}\frac{d\dot{x}}{dx}&=\underset{x\rightarrow 0}{\lim} \gamma x^{\gamma-1}(1+\alpha D)-(1-\alpha D)\\
\underset{x \rightarrow 1}{\lim}\frac{d\dot{x}}{dx}&=\underset{x \rightarrow 1}{\lim}\gamma(1-x)^{\gamma-1}(1-\alpha D)-(1+\alpha D).
\end{split}
\end{equation*}

Thus, if $\gamma < 1$, $x=0$ is stable if $\lim_{x\rightarrow 0}(1+\alpha D)<0$ and $x=1$ is stable if $\lim_{x\rightarrow 0}(1-\alpha D)<0$. If $\gamma > 1$, $x=0$ is stable if $\lim_{x\rightarrow 1}-(1-\alpha D)<0$ and $x=1$ is stable if $\lim_{x\rightarrow 1}-(1+\alpha D)<0$. Notice that this is equivalent to checking the sign of $\bar{D}$ at each endpoint. Specifically, if $D$ is bounded on $x\in[0,1]$ and $\gamma<1$, then

\begin{equation*}
\begin{split}
\underset{x\rightarrow 0}{\lim}\bar{D}&=\underset{x\rightarrow 0}{\lim}x^{\gamma-1}(1+\alpha D)-(1-\alpha D)\\
\underset{x\rightarrow 1}{\lim}\bar{D}&=\underset{x\rightarrow 1}{\lim}(1+\alpha D)-(1-x)^{\gamma-1}(1-\alpha D)
\end{split}
\end{equation*}
which implies that

\begin{equation*}
\begin{split}
\textrm{sign}\left(\underset{x\rightarrow 0}{\lim}\frac{d\dot{x}}{dx}\right)&=\textrm{sign}\left(\underset{x\rightarrow 0}{\lim}\bar{D}\right)=\textrm{sign} \left(1+\alpha D|_{x=0}\right)\\
\textrm{sign}\left(\underset{x\rightarrow 1}{\lim}\frac{d\dot{x}}{dx}\right)&=\textrm{sign}\left(\underset{x\rightarrow 1}{\lim}\bar{D}\right)=-\textrm{sign} \left(1-\alpha D|_{x=1}\right).
\end{split}
\end{equation*}

On the other hand, if $\gamma > 1$,

\begin{equation*}
\begin{split}
\underset{x\rightarrow 0}{\lim}\bar{D}&=-(1-\alpha D)\\
\underset{x\rightarrow 1}{\lim}\bar{D}&=1+\alpha D
\end{split}
\end{equation*}
which implies that

\begin{equation*}
\begin{split}
\textrm{sign}\left(\underset{x\rightarrow 0}{\lim}\frac{d\dot{x}}{dx}\right)&=\textrm{sign}\left(\underset{x\rightarrow 0}{\lim}\bar{D}\right)=-\textrm{sign}\left(1-\alpha D|_{x=0}\right)\\
\textrm{sign}\left(\underset{x\rightarrow 1}{\lim}\frac{d\dot{x}}{dx}\right)&=\textrm{sign}\left(\underset{x\rightarrow 1}{\lim}\bar{D}\right)=\textrm{sign}(1+\alpha D|_{x=1}).
\end{split}
\end{equation*}

Note that, even though the expression for $d\dot{x}/dx$ can be complicated, especially for internal fixed points, by  continuity of $\bar{D}$ we know that there is at least one unstable fixed point in $(0,1)$ (because $\bar{D}$ is up-crossing the x-axis) when the fixed points at both ends are are stable; and there is at least one stable fixed point in $(0,1)$ (because $\bar{D}$ is down-crossing the x-axis) when the fixed points at both ends are unstable.

\subsection{Linear games}\label{sec:linear games}

Building on the analyses in Section~\ref{sec:fixed points and stability}, we can obtain more concrete results for general linear games. The analysis in this Section, in turn, can be directly applied to specific linear games to produce the results shown in Figure~\ref{tab:linear games}.

The payoff difference of an arbitrary linear $2 \times 2$ game with payoff matrix {\tiny$\setlength\arraycolsep{1pt}\begin{pmatrix} R & S \\ T & P \end{pmatrix}$} can be written as $D(x)=(P+R-S-T)x+S-P$. For such a game we can obtain an explicit expression for the internal fixed points when $\gamma=2$ or $\gamma=1/2$, but the expression may be complicated. More, generally, we can often narrow down the {\it{number}} of internal fixed points depending on the game's payoff matrix and value of complexity parameter $\gamma$.

The equation we must solve to obtain the fixed point becomes 
\begin{equation*}
\left(\frac{x}{1-x}\right)^{\gamma-1}=\frac{1-\alpha[(P+R-S-T)x+S-P]}{1+\alpha[(P+R-S-T)x+S-P]}=\frac{2}{\alpha [(P+R-S-T)x+S-P]+1}-1.
\end{equation*}

Observe that for $\gamma > 1$, the left-hand side (LHS) is monotonic increasing. Conversely, for $\gamma < 1$, the LHS is monotonic decreasing. This yields three noteworthy scenarios: Firstly, in the case where $P+R-S-T=0$ (characteristic of the Donation game), the right-hand side (RHS) remains constant, implying a maximum of one internal fixed point regardless of $\gamma$'s sign. Given that the LHS is strictly positive in the interior $x\in(0,1)$, the RHS must also be positive for a solution to exist. This condition necessitates that either $S-P<0$ and $\alpha<1/(P-S)$, or $0<S-P$ and $\alpha<1/(S-P)$ for an internal fixed point to be feasible; Secondly, when $P+R-S-T > 0$ (typical of the Coordination game), the RHS is monotonically decreasing. Therefore, there can be at most one internal fixed point if $\gamma > 1$, with the LHS increasing monotonically; Finally, if $P+R-S-T < 0$ (it can be true, for example, in Prisoner's dilemma and Snowdrift game depending on the parameters), the RHS increases monotonically. In this scenario, there can be at most one internal fixed point is possible if $\gamma < 1$, corresponding with a monotonically decreasing LHS.

Regarding the stability of fixed points, we have

\begin{equation*}
\begin{split}
\textrm{sign}(\left.1+\alpha D\right|_{x=0})=\textrm{sign}(1+\alpha(S-P))\\
-\textrm{sign}(\left.1-\alpha D\right|_{x=1})=-\textrm{sign}(1-\alpha(R-T))\\
-\textrm{sign}(\left.1-\alpha D\right|_{x=0})=-\textrm{sign}(1-\alpha(S-P))\\
\textrm{sign}(\left.1+\alpha D\right|_{x=1})=\textrm{sign}(1+\alpha(R-T)).
\end{split}
\end{equation*}
Therefore, for $\gamma<1$

\begin{equation*}
\begin{split}
x=0\textrm{ is }
&\left\{
\begin{array}{ll}
\textrm{unstable iff} & P-S < \frac{1}{\alpha} \\
\textrm{stable iff} & \frac{1}{\alpha} < P-S
\end{array}
\right.\\
x=1\textrm{ is }
&\left\{
\begin{array}{ll}
\textrm{unstable iff }& -\frac{1}{\alpha} < T-R \\
\textrm{stable iff }&  T-R < -\frac{1}{\alpha}
\end{array}
\right.
\end{split}
\end{equation*}

There is a stable internal equilibrium if both $P-S<1/\alpha$ and $-1/\alpha<T-R$ are true, and there is an unstable internal equilibrium if both $1/\alpha<P-S$ and $T-R<-1/\alpha$ are true. 

For $\gamma>1$

\begin{equation*}
\begin{split}
x=0 \textrm{ is }
&\left\{
\begin{array}{ll}
\textrm{stable iff}& -\frac{1}{\alpha} < P-S \\
\textrm{unstable iff}&  P-S < -\frac{1}{\alpha}
\end{array}
\right.\\
x=1 \textrm{ is }
&\left\{
\begin{array}{ll}
\textrm{stable iff}& T-R < \frac{1}{\alpha} \\
\textrm{unstable iff}&  \frac{1}{\alpha} < T-R.
\end{array}
\right.
\end{split}
\end{equation*}

There is an unstable internal equilibrium if both $-1/\alpha<P-S$ and $T-R<1/\alpha$ are true, and there is a stable internal if both $P-S<-1/\alpha$ and $1/\alpha<T-R$ are true.

\subsection{Generalization to games with more than two strategies}

We denote the probability of the number of type $X_{j}$ individuals increasing or decreasing by one at state $\vec{x}$ by $T^{\pm}_{X_{j}}(\vec{x})$. In a well-mixed population, this formulation of $\rho$ results in dynamics that can be described by the following transition probabilities.

\begin{equation*}
\begin{split}
T^{+}_{X_{j}}(\vec{x})&=\sum_{k}x_{k}\sum_{\vec{v}\in \mathbb{Z}_{+}^{m}}\left.\rho^{X_{k}\to X_{j}}\right|_{M,\vec{n}}P(\vec{n}=\vec{v})\\
&=\sum_{k}x_{k}\sum_{\vec{v}\in \mathbb{Z}_{+}^{m}}\frac{(n_{X_{j}}/n)^{\gamma}}{1+\sum_{l}\exp\left[s\left(f_{X_{l}}(\vec{n}/n)-f_{X_{j}}(\vec{n}/n)\right)\right]}P(\vec{n}=\vec{v})\\
T^{-}_{X_{j}}(\vec{x})&=x_{j}\sum_{k}\sum_{\vec{v}\in \mathbb{Z}_{+}^{m}}\left.\rho^{X_{j}\to X_{k}}\right|_{M,\vec{n}}P(\vec{n}=\vec{v})\\
&=x_{j}\sum_{k}\sum_{\vec{v}\in \mathbb{Z}_{+}^{m}}\frac{(n_{X_{k}}/n)^{\gamma}}{1+\sum_{l}\exp\left[s\left(f_{X_{l}}(\vec{n}/n)-f_{X_{k}}(\vec{n}/n)\right)\right]}P(\vec{n}=\vec{v})
\end{split}
\end{equation*}

where $P(\vec{n}=\vec{v})$ is a multinomial distribution:

\begin{equation*}
\begin{split}
P(\vec{n}=\vec{v})&=f(n_{X_{1}},\cdots,n_{X_{m}};n;x_{1},\cdots,x_{m})\\
&=
\left\{
\begin{array}{ll}
\frac{n!}{n_{X_{1}}!\cdots n_{X_{m}}!}x_{1}^{n_{X_{1}}}\times \cdots \times x_{m}^{n_{X_{m}}} & \textrm{if }\sum_{i=1}^{m}n_{X_{i}}=n \\
0 & \textrm{otherwise}.
\end{array}
\right.
\end{split}
\end{equation*}

Denote the probability that the system is in state $\vec{x}$ at time $\tau$ by $P^{\tau}(\vec{x})$. We then have

\begin{equation*}
\begin{split}
P^{\tau+1}(\vec{x})-P^{\tau}(\vec{x})=&\sum_{i}P^{\tau}(x_{1},\cdots,x_{i}-N^{-1},\cdots,x_{m-1})T^{+}_{X_{i}}(x_{1},\cdots,x_{i}-N^{-1},\cdots,x_{m-1})\\
&+\sum_{i}P^{\tau}(x_{1},\cdots,x_{i}+N^{-1},\cdots,x_{m-1})T^{-}_{X_{i}}(x_{1},\cdots,x_{i}-N^{-1},\cdots,x_{m-1})\\
&-\sum_{i}P^{\tau}(\vec{x})T^{+}_{X_{i}}(\vec{x})\\
&-\sum_{i}P^{\tau}(\vec{x})T^{-}_{X_{i}}(\vec{x}).
\end{split}
\end{equation*}

We define $p(x_{1},\cdots,x_{m-1},t) \triangleq NP^{\tau}(x_{1},\cdots,x_{m-1})$. Rewriting the above expression in terms of $p$'s and applying Taylor expansion, we obtain

\begin{equation*}
\begin{split}
\frac{dp(\vec{x},t)}{dt}=&-\sum_{i}\frac{\partial}{\partial x_{i}}\left[\left(T^{+}_{X_{i}}(\vec{x})-T^{-}_{X_{i}}(\vec{x})\right)p^{\tau}(\vec{x},t)\right]\\
&+\sum_{i}\frac{1}{2}\frac{\partial^{2}}{\partial x_{i}^{2}}\left[\frac{T^{+}_{X_{i}}(\vec{x})+T^{-}_{X_{i}}(\vec{x})}{N}p^{\tau}(\vec{x},t)\right]
\end{split}
\end{equation*}
which has an equivalent Langevin equation
$dX=A(x,t)dt+B(x,t)dW_{t}$
where $W$ is now an $m-1$ dimensional Wiener process and 

\begin{equation*}
\begin{split}
A&=
\begin{pmatrix}
T^{+}_{X_{1}}(\vec{x})-T^{-}_{X_{1}}(\vec{x})\\
\vdots\\
T^{+}_{X_{m-1}}(\vec{x})-T^{-}_{X_{m-1}}(\vec{x})
\end{pmatrix}
\\
B &= 
\begin{pmatrix}
\frac{T^{+}_{X_{1}}(\vec{x}) - T^{-}_{X_{1}}(\vec{x})}{N} & 0 & \cdots & 0 \\
0 & \frac{T^{+}_{X_{2}}(\vec{x}) - T^{-}_{X_{2}}(\vec{x})}{N} & \ddots & \vdots \\
\vdots & \ddots & \ddots & 0 \\
0 & \cdots & 0 & \frac{T^{+}_{X_{m}}(\vec{x}) - T^{-}_{X_{m}}(\vec{x})}{N}
\end{pmatrix}.
\end{split}
\end{equation*}
Taking the limit of large $N$ the noise terms become negligible, which leads to the system of ODEs:

\begin{equation*}
\dot{x}_{j}=T^{+}_{X_{j}}(\vec{x})-T^{-}_{X_{j}}(\vec{x}).
\end{equation*}

We further apply the following theorem regarding multivariate Bernstein polynomials on a simplex \cite{bayad2011} to obtain an approximation that is valid for sufficiently large $n$.

\begin{theorem*}
Let $g:\Delta_{k}\rightarrow \mathbb{R}$ be a continuous function where $\Delta_{k}$ is a k-dimensional simplex defined as $\Delta_{k}=\{\vec{p}=(p_{1},\cdots,p_{k})\in [0,1]^{k}:p_{1}+\cdots+p_{k}\leq 1\}$. Let
\begin{equation*}
B_{g,t,k}(\vec{p})=\sum_{|\vec{v}|\leq t}f\left(\frac{\vec{v}}{t}\right)\frac{v!}{\vec{v}!(t-|\vec{v}|)!}\vec{p}^{\vec{v}}(1-|\vec{p}|)^{t-|\vec{v}|}
\end{equation*}
where $\vec{v}=(v_{1},\cdots,v_{k})\in \mathbb{N}^{k}$, $t\in \mathbb{N}\backslash 0$, $|\vec{v}|=v_{1}+\cdots+v_{k}$, $\vec{v}!=v_{1}!\cdots v_{k}!$, $\vec{p}^{\vec{v}}=p_{1}^{v_{1}}\cdots p_{k}^{v_{k}}$, $|\vec{p}|=p_{1}+\cdots+p_{k}$. Then, $B_{g,n,k}(\vec{p})$ converges uniformly to $f(\vec{p})$ as $v\rightarrow \infty$.
\end{theorem*}

We apply this theorem by letting $g=\rho^{X_{k}\to X_{j}},\rho^{X_{j}\to X_{k}}$, $k=m-1$, $\vec{p}=\vec{x}=(x_{1},\cdots,x_{m-1})$, and $t=n$. By this theorem as well as subsequent application of a Taylor expansion around $s=0$, we obtain the following approximation:

\begin{equation*}
\begin{split}
T^{+}_{X_{j}}(\vec{x})&\approx\sum_{k}x_{k}x_{j}^{\gamma}\left[\frac{1}{m}-\frac{s}{m^{2}}\sum_{l\neq j}\left(f_{X_{l}}(\vec{x})-f_{X_{j}}(\vec{x})\right)\right]\\
T^{-}_{X_{j}}(\vec{x})&\approx x_{j}\sum_{k}x_{k}^{\gamma}\left[\frac{1}{m}-\frac{s}{m^{2}}\sum_{l \neq k}\left(f_{X_{l}}(\vec{x})-f_{X_{k}}(\vec{x})\right)\right]
\end{split}
\end{equation*}
where $\vec{x}=(x_{1},\cdots,x_{m})$. Substituting the approximation into the approximate dynamics and setting $\gamma=1$, the system of ODE reduces to the standard replicator equation:

\begin{equation*}
\begin{split}
\dot{x}_{j}&=\sum_{k}x_{k}x_{j}\left[\frac{1}{m}-\frac{s}{m^{2}}\sum_{l\neq j}\left(f_{X_{l}}(\vec{x})-f_{X_{j}}(\vec{x})\right)\right]-x_{j}\sum_{k}x_{k}\left[\frac{1}{m}-\frac{s}{m^{2}}\sum_{l\neq k}\left(f_{X_{l}}(\vec{x})-f_{X_{k}}(\vec{x})\right)\right]\\
&=x_{j}\sum_{k}x_{k}\frac{s}{m^{2}}\sum_{l\neq k}\left(f_{X_{l}}(\vec{x})-f_{X_{k}}(\vec{x})\right)-\sum_{k}x_{k}x_{j}\frac{s}{m^{2}}\sum_{l\neq j}\left(f_{X_{l}}(\vec{x})-f_{X_{j}}(\vec{x})\right)\\
&=\frac{sx_{j}}{m^{2}}\sum_{k}x_{k}\left[\sum_{l\neq k}\left(f_{X_{l}}(\vec{x})-f_{X_{k}}(\vec{x})\right)-\sum_{l\neq j}\left(f_{X_{l}}(\vec{x})-f_{X_{j}}(\vec{x})\right)\right]\\
&=\frac{sx_{j}}{m^{2}}\sum_{k}x_{k}\left[\sum_{l}f_{X_{l}}(\vec{x})-f_{X_{k}}(\vec{x})-(m-1)f_{X_{k}}(\vec{x})-\sum_{l}f_{X_{l}}(\vec{x})+f_{X_{j}}(\vec{x})+(m-1)f_{X_{j}}(\vec{x})\right]\\
&=\frac{sx_{j}}{m^{2}}\sum_{k}x_{k}\left[m\left(f_{X_{j}}(\vec{x})-f_{X_{k}}(\vec{x})\right)\right]\\
&=\frac{sx_{j}}{m}\left[f_{X_{j}}(\vec{x})-\sum_{k}x_{k}f_{X_{k}}(\vec{x})\right]
\end{split}
\end{equation*}
where the last equality follows from the fact that $\sum_{k}x_{k}=1$.

\subsection{An alternative construction of the process for multi-choice games}

There is an alternative mechanism, parameterized by $\kappa \in \mathbb{Z}_{+}$, that gives rise to the standard replicator dynamics when $\kappa=1$. Suppose a focal individual $i$ of type $X_{j}$ samples $\kappa$ opponents and considers switching to another type only if all opponents it interacts with are of that type. If all the opponents are of that type, the transition occurs with some probability that depends on the difference in the payoff between the focal individual's original type and the opponents' type. Specifically, the focal individual switches to that type with probability

\begin{equation*}
\rho^{X_{j}\to X_{k}}=
\left\{
\begin{array}{ll}
\frac{1}{1+\exp \left[s\left(f_{X_{j}}(\vec{n})-f_{X_{k}}(\vec{n})\right)\right]} & \text{if all $\kappa$ opponents are of type $k$} \\
0 & \text{otherwise}.
\end{array}
\right.
\end{equation*}
We then clearly have

\begin{equation*}
\rho^{X_{j}\to X_{k}}=\frac{x^{\kappa}_{k}}{1+\exp \left[s\left(f_{X_{j}}(\vec{n})-f_{X_{k}}(\vec{n})\right)\right]}.
\end{equation*}

The dynamics can be described by transition probabilities:

\begin{equation*}
\begin{split}
T^{+}_{X_{j}}(\vec{x})&=\sum_{k\neq j}x_{k}\sum_{\vec{v}\in \mathbb{Z}_{+}^{m}}\left.\rho^{X_{k}\to X_{j}}\right|_{M,\vec{n}}P(\vec{n}=\vec{v})\\
T^{-}_{X_{j}}(\vec{x})&=x_{j}\sum_{k\neq j}\sum_{\vec{v}\in \mathbb{Z}_{+}^{m}}\left.\rho^{X_{j}\to X_{k}}\right|_{M,\vec{n}}P(\vec{n}=\vec{v})
\end{split}
\end{equation*}
where $P(\vec{n}=\vec{v})$ is a multinomial distribution. By the same procedure as above, we obtain a system of ODEs that approximate the dynamics when $N$ and $n$ are large.

\begin{equation*}
\dot{x}_{j}=x^{\kappa}_{j}\sum_{k\neq j}\frac{x_{k}}{1+\exp \left[s\left(f_{X_{j}}(\vec{x})-f_{X_{k}}(\vec{x})\right)\right]}-x_{j}\sum_{k\neq j}\frac{x^{\kappa}_{k}}{1+\exp\left[s\left(f_{X_{k}}(\vec{x})-f_{X_{j}}(\vec{x})\right)\right]}.
\end{equation*}

By Taylor expansion, we obtain a further approximation.

\begin{equation*}
\dot{x}_{j}=x_{j}^{\kappa}\sum_{k\neq j}x_{k}\left[\frac{1}{2}-\frac{s}{4}\left(f_{X_{k}}(\vec{x})-f_{X_{j}}(\vec{x})\right)\right]-x_{j}\sum_{k\neq j}x^{\kappa}_{k}\left[\frac{1}{2}-\frac{s}{4}\left(f_{X_{j}}(\vec{x})-f_{X_{k}}(\vec{x})\right)\right].
\end{equation*}

By setting $\kappa=1$, this reduces the replicator equation:
\begin{equation*}
\dot{x}_{j}=\frac{s}{4}x_{j}\left[f_{X_{j}}(\vec{x})-\sum_{k}x_{k}f_{X_{k}}(\vec{x})\right].
\end{equation*}

The dynamics exhibited by this version are identical to our main model when $m=2$ and it appears to remain similar when $m=3$ for simple games (Supplementary Material).
}

\printbibliography

\newpage

\renewcommand{\thefigure}{S\arabic{figure}}
\renewcommand{\thetable}{S\arabic{table}}
\renewcommand{\thesubsection}{S\arabic{subsection}}
\setcounter{figure}{0} 
\setcounter{table}{0}
\setcounter{subsection}{0}

\section*{Supplementary Material}

\subsection{Monte Carlo simulation results}

Table~\ref{tab:simulation} shows the strategic frequencies at the quasi-stationary state, which were computed by running simulations for $10^{6}$ generations and taking the average of the frequencies over the last $10^{5}$ generations. Corresponding predictions derived from the infinite-population ODE are indicated by brackets. The quasi-stationary distribution in the simulations correspond well with the stable fixed points of the continuous dynamics.
\begin{table}[!ht]
\centering
\begin{tabular}{cccccc}
\toprule
 &  & \multicolumn{4}{c}{Initial frequency of strategy 1} \\
\cmidrule(rl){3-6}
Game &\textbf{$\gamma$} & 0.2 & 0.4 & 0.6 & 0.8 \\
\midrule
\multirow{3}{*}{Donation} & 1 & 0 (0) & 0 (0) & 0 (0) & 0 (0) \\
 & 2 & 0 (0) & 0 (0) & 0 (0) & 1 (1) \\
 & 0.5 & 0.235 (0.225) & 0.225 (0.225) & 0.222 (0.225) & 0.228 (0.225) \\
\midrule
\multirow{3}{*}{Snowdrift} & 1 & 0.771 (0.75) & 0.745 (0.75) & 0.757 (0.75) & 0.762 (0.75) \\
 & 2 & 0 (0) & 1 (1) & 1 (1) & 1 (1) \\
 & 0.5 & 0.590 (0.593) & 0.590 (0.593) & 0.593 (0.593) & 0.592 (0.593) \\
\midrule
\multirow{3}{*}{Coordination} & 1 & 0 (0) & 0 (0) & 1 (1) & 1 (1) \\
 & 2 & 0 (0) & 0 (0) & 1 (1) & 1 (1) \\
 & 0.5 & 0.211 (0.149) & 0.898 (0.941) & 0.900 (0.941) & 0.904 (0.941) \\
\bottomrule
\end{tabular}
\caption{\small \textbf{Monte Carlo simulations of the discrete stochastic process.} The quasi-stationary distribution of the discrete time/state stochastic process correspond with the stable equilibria of the continuous-time ODE approximation. In these simulation, a focal player interacts is a sub-population of size $n=500$, which is  substantially smaller than the total size of the population ($N=2500$). These results confirm the accuracy of the ODE approximation  under realistic, finite-population parameters. To maintain parity with Figure 1, the same game matrices are used, and the strength of payoff-biased selection is  $s=2,\alpha=0.3$, which accounts for the re-scaling in the derivation of the continuous limit ODE (see Methods).}
\label{tab:simulation}
\end{table}

\subsection{Three-strategy Prisoner's Dilemma}

The payoff matrix of the three-strategy prisoner's Dilemma is given by
\begin{equation*}
\begin{pmatrix}
b_{1}-c_{1}&b_{2}-c_{1}&-c_{1}\\
b_{1}-c_{2}&b_{2}-c_{2}&-c_{2}\\
b_{1}&b_{2}&0
\end{pmatrix}
\label{eq:3-strategy PD}
\end{equation*}
where $b_{2}>b_{1},c_{2}>c_{1}$ and $b_{1}-c_{1}>b_{2}-c_{2}$.  The first strategy, full cooperation, is more efficient than the second strategy, moderate cooperation, in the sense that the payoff of the entire population is higher when all individuals adopt the former strategy than when they adopt the second strategy. Also, moderate cooperation is more individually rational than full cooperation as adopting the former strategy always gain a higher payoff. The least efficient monomorphic population consists of defectors, but it is individually rational to defect.

\begin{figure}[!ht]
\centering
\includegraphics[width=\linewidth]{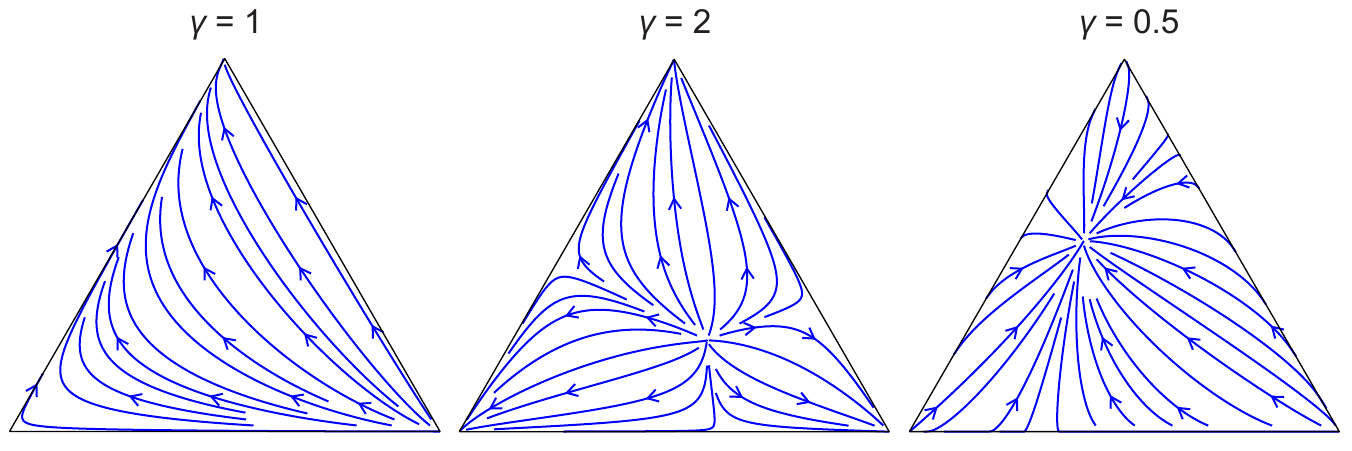}
\caption{\small \textbf{Three-strategy Prisoner's Dilemma} 
Each vertex represents the monomorphic population of each strategy. The right and left bottom verteces represent full and moderate cooperators, and the top vertex represents defectors. When $\gamma=1$ (classic replicator equation), the dynamics exhibits a cyclic nature. It is attracted to each monomorphic population when $\gamma>1$ and attracted to the mixed strategy population  when $\gamma<1$. Parameters: $s=3,b_{1}=3,b_{2}=4,c_{1}=1,c_{2}=3$.}
\label{fig:3-strategy PD}
\end{figure}

Under standard replicator dynamics ($\gamma=1$), the dynamics leads to the all-defector state. Whereas for $\gamma > 1$, the population is attracted toward each monomorphic population state, and the strategy that is more individually rational strategy has a larger basin of attraction. When $\gamma < 1$, the population is attracted to a mixed equilibrium with the more individually rational strategy having a higher frequency (Figure~\ref{fig:3-strategy PD}).

\subsection{The dynamics of the alternative construction}

\begin{figure}[!ht]
\centering
\includegraphics[width=0.68\textwidth]{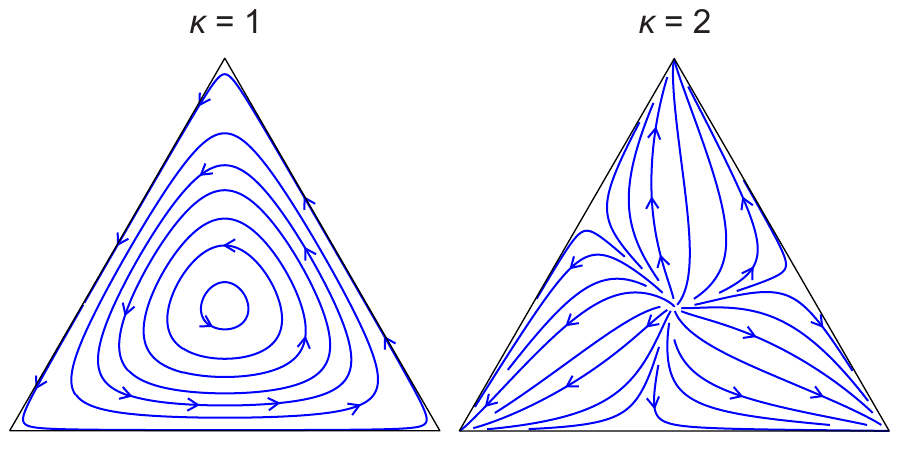}
\caption{\small \textbf{Rock-Paper-Scissor game under the alternative dynamics} 
Each vertex represents the monomorphic population of each strategy. When $\kappa=1$ (classic replicator equation), the dynamics exhibits a cyclic nature. It is attracted to each monomorphic population when $\kappa>1$. Parameter: $s=0.3$.}
\label{fig:RPS alt}
\end{figure}

\begin{figure}[!ht]
\centering
\includegraphics[width=0.68\linewidth]{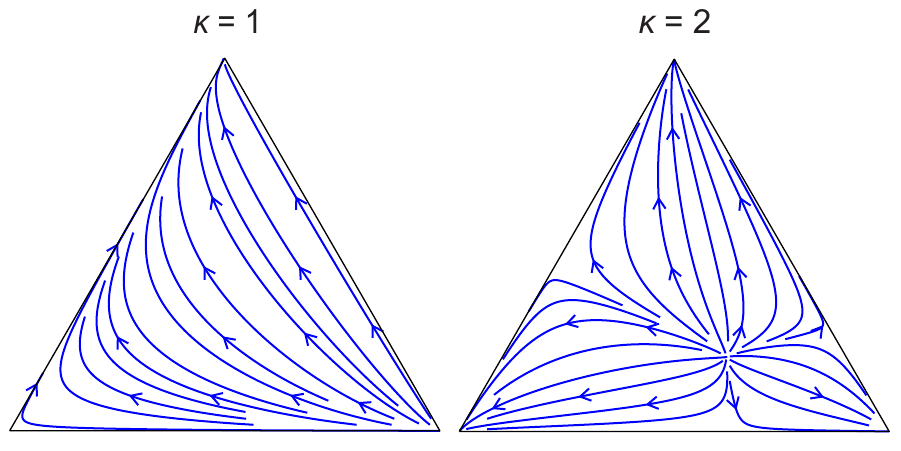}
\caption{\small \textbf{Three-strategy Prisoner's Dilemma under the alternative dynamics} 
Each vertex represents the monomorphic population of each strategy. The right and left bottom verteces represent full and moderate cooperators, and the top vertex represents defectors. When $\kappa=1$ (replicator equation), the dynamics is attracted toward the all-defector population. It is attracted to each monomorphic population when $\kappa>1$ with different sizes of attraction, largest for the all-defector population and smallest for the all-full-cooporator population. Parameters are the same as in Figure~\ref{fig:3-strategy PD}.}
\label{fig:3 strategy PD alt}
\end{figure}

The dynamics exhibited by the continuous limit ODE derived from the alternative construction is identical to our main model when $m=2$, and it appears to remain similar when $m=3$ for some typical linear games as shown in Figures~\ref{fig:RPS alt} and~\ref{fig:3 strategy PD alt}.

\end{document}